# Assessment of toxic metals and hazardous substances in tattoo inks using Sy-XRF, AAS and Raman spectroscopy

Marta Manso[1*], Sofia Pessanha[1], Mauro Guerra[1], Uwe Reinholz[2], Cláudia Afonso[3], Martin Radtke[2], Helena Lourenço[3], Maria Luísa Carvalho[1] and Ana Guilherme Buzanich[2]

[1]LIBPhys-UNL Laboratório de Instrumentação, Engenharia Biomédica e Física da Radiação, Departamento de Física, Faculdade de Ciências e Tecnologia (FCT), Universidade Nova de Lisboa, 2829-516 Caparica, Portugal.
[2]Bundesanstalt fuer Materialforschung und pruefung (BAM), Richard-Willstaetter-Strasse 11, 12489 Berlin, Germany.
[3]IPMA Portuguese Institute for the Sea and Atmosphere, Division of Aquaculture and Upgrading, Avenida Brasılia, 1449-006 Lisboa, Portugal.

*Telephone number: +351 917408908   e-mail: marta.sampaio@fct.unl.pt

## Abstract

Synchrotron radiation X-Ray Fluorescence spectroscopy, in conjunction with Atomic Absorption and Raman spectroscopy, were used to analyse a set of top brand tattoo inks to investigate the presence of toxic elements and hazardous substances. The Cr, Cu and Pb contents were found to be above the maximum allowed levels established by the Council of Europe through the resolution ResAP(2008)1 on requirements and criteria for the safety of tattoos and permanent make-up. Raman analysis has revealed the presence of a set of prohibited substances mentioned in ResAP(2008)1, among which are the pigments Blue 15, Green 7 and Violet 23. Other pigments that were identified in white, black, red, and yellow inks, are the Pigment White 6, Carbon Black, Pigment Red 8 and a diazo yellow, respectively. The present results show the importance of regulating tattoo ink composition.

## Introduction

Tatooing is an artistic and social expression that consists of permanently marking the skin by an intradermal injection of pigments, lakes or dyes. Other ingredients such as solvents, pH regulators, emollients and thickeners are also added to the colourants as a way to enhance or stabilize the colourants [1]. Although this practice can be traced back to the Neolithic [2], it has seen a rise all over the globe, representing now a very important socio-cultural phenomenon. For example, in the United States, up to 24% of the population is tattooed, whereas in European countries one can find tattoos in only 12% of the population [3]. There are increasing reports of adverse reactions such as allergies and granulomas which appear immediately or years after tattooing [4], and that have been related to the composition of the inks [5-6].

Manufacturers as well as suppliers are not usually obliged to disclose the components of their tattoo inks, as a way to maintain proprietary knowledge of their formulations [7]. Also, the pigments that are used in the inks are not always produced with the purpose of being delivered in Human dermis for long periods of time, which allow that no risk assessment is performed before commercialization. Hence, in general tattoo inks are not officially controlled by any public organism, which means that tattoo inks may contain harmful components such as chromium VI in Cr oxides; Ni, Cr, Cu and Co in iron oxides; aromatic amines in azo-colorants as well as polycyclic aromatic hydrocarbons in carbon black [3].

In 2008, the Council of Europe (CoE) adopted a resolution regulating the criteria for the safety of tattoos and Permanent Make-Up (PMU) [1]. In this document the CoE recommended that the governments of member states follow some good practices, namely to limit the concentration of impurities in products to be used in tattoos. Furthermore, the use of certain organic substances in tattoo inks are completely forbidden due to their carcinogenic, mutagenic, or other adverse properties. Nevertheless, analytical studies of ink composition show that these propositions are not always acquiesced.

An Italian survey of 56 inks [8] showed, through the use of Sector Field Inductively Coupled Plasma Mass Spectrometry (SF-ICP-MS), that several toxic elements such as V, Mn, Cd, Sb and Pb had, in some cases, concentrations higher than 1 µg/g. In the same survey, the allergenic metals, Cr, Ni and Co presented concentrations well above the safe limit in 62.5%, 16.1% and 1.8% of the cases, respectively. In another study in Iran [9], Eghbali *et al.*, found that some of the commercially available tattoo inks had Cd mass fractions as high as 2.1 µg/g while the permitted limit is only 0.2 µg/g. The presence of a forbidden substance, Pb, was also found among the same set of inks and usually associated to the black coloured inks. Schreiver *et al.* [10] have used pyrolysis coupled to online gas chromatography and electron impact ionization mass spectrometry, to identify both purified pigments and tattoo inks in their final form, having built for this purpose a pyrogram library. Poon *et al.* [11] demonstrated the possibility of using Raman spectroscopy for organic pigments, by analysing sections of cryogenically cooled pig skin samples upon which tattoos were imprinted. Raman spectroscopy was also used to identify characteristic coal Raman bands in skin biopsies from tattooed coal-miners [12].

Recently, due to the increasing concern on metallic nanomaterials and their health risks, Bocca *et al.* [13] used a set of analytical techniques such as transmission electron microscopy, dynamic light scattering, single particle ICP-MS as well as asymmetric flow field fractionation with a multi-angle light scattering detector, to characterize the size of nano and micro particles in tattoo inks. Høgsberg *et al.* [14] have also measured the particle sizes by laser diffraction, electron microscopy and X-ray diffraction, for a better evaluation of human exposure to these metallic particles.

In this work, our goal is to provide information about the elemental content and molecular nature of the materials used in a set of tattoo inks, evaluating if it complies with the Resolution ResAP(2008)1 on its safety requirements. A secondary aim is to better characterize the inks, filling the missing scientific information on its composition. For this purpose, a set of a top brand tattoo inks was firstly analysed by synchrotron radiation X-ray fluorescence (Sy-XRF) spectroscopy enabling the identification of potentially toxic elements whose content was then measured by Atomic Absorption Spectrometry (AAS). As the compound toxicity is not just related to the mass fraction values but also to the bioavailability, which depends on the chemical species, Raman spectroscopy was used for the identification of molecular compounds within the ink.

## Materials and Methods

The set of tattoo inks that was analysed in this work is from the Kuro Sumi brand. The colour labels are, according to the company assignation, Tsunami blue, Black, White, Violent violet, Tomato red, Forest is green and Deep yellow.

Synchrotron Radiation X-Ray Fluorescence Spectrometry (Sy-XRF)

Sy-XRF analysis was performed at the BAM*line* at the synchrotron facility BESSY II in Berlin, Germany [15]. Due to expected mass fractions of most elements in the ppm range, the use of Sy-XRF was mandatory to obtain a good signal to noise ratio. The samples were prepared by placing a few drops of the ink in a kapton foil and let them dry in air. The samples were then placed at a 45º angle relative to both the incoming beam and the detector, resulting in a 90º angle between the incident and scattered photons in order to minimize Compton and Rayleigh scattering [16]. An excitation energy of 20 KeV, monochromatized by a double multilayer monochromator, was used throughout all of the measurements. Regarding spectra acquisition, a silicon drift detector, XFlash® 5030 (Bruker Nano GmbH, Berlin, Germany), was used, featuring an energy resolution of 127 eV@5.9 KeV, and the acquisition time for each sample was 1000 s.

Atomic Absorption Spectrometry (AAS)

The sample preparation for this technique required that about 0.2 g of each ink were weighed and then digested in a microwave oven MARS 5 (CEM, Matthews, North Carolina, USA), with a mixture of nitric acid (65%, w/w) and hydrogen peroxide (30%, w/w). The resulting solution was then diluted with water. The detection of elements Cd, Cr, Cu, Ni and Pb was performed according to the European norm NP EN 14084 (2003) [17], with Cd and Pb being quantified using the graphite furnace AAS method with Spectr 220Z apparatus (Varian, Palo Alto, California, USA) with a Zeeman correction ($\lambda$ = 283.3 and $\lambda$ = 228.8 nm, respectively for Pb e Cd). Regarding Cr, Cu and Ni, their concentration was measured with AAS with flame atomisation in a Varian apparatus Spectr AA 55B (Varian, Palo Alto, USA) with a background deuterium correction ($\lambda$ = 357.9 nm, $\lambda$ = 324.8 nm, $\lambda$ = 232.0, respectively for Cr, Cu e Ni). Total Hg content was determined by AAS using an automatic Hg analyser AMA 254 (Leco corporation, Saint Joseph, Michigan, USA) ($\lambda$ = 253.7 nm) [18].

μ-Raman Spectroscopy

Micro-Raman spectroscopy was performed directly on samples that were left to dry in a glass substrate. The apparatus is an XploRA confocal spectrometer (Horiba-Jobin Yvon, France), which was operated at a wavelength of 785 nm, using a 100× magnification objective with a pinhole of 300 μm and an entrance slit of 100 μm. The scattered light from the sample was then collected by the same objective and dispersed by a 1200 lines/mm grating onto an air-cooled CCD detector iDus (Andor Technology Ltd., Belfast, UK). The analysis was performed in the 100-2000 cm$^{-1}$ range and the spectral deconvolution was made with the Spectral ID software from Horiba using its built-in database as well as with other published Raman spectra.

Results and Discussion

The results obtained in this work by Sy-XRF show variations in the metallic composition of the inks although, as a general trend, Ca, Fe, Cu and Sr were present in all the samples (see Fig. 1). The white, blue and violet inks spectra show very intense Ti characteristic peaks that should be related to the use of $TiO_2$ either as a carrier for dyes or as a pigment. Nevertheless, the Raman characteristic bands of $TiO_2$ were only identified at 141, 228, 448 and 611 cm$^{-1}$ in the white ink and in the form of rutile indicating the use of Pigment White 6, similarly to what has been found in reference [11]. The presence of these bands in the other inks could have been hindered by the Raman bands of copper phthalocyanine and dioxazine dyes present in blue and violet inks, respectively. Bismuth was also found in the black ink, which can be due to the fact that bismuth oxychloride is often used as an effect pigment in cosmetics, plastics and paints. This effect pigment can have its ultraviolet stability or weather fastness properties increased if coated with materials as 2-hydroxy benzophenones and rare earths [19], although in this work we haven't seen neither rare earths in Sy-XRF or AAS, neither 2-hydroxy benzophenones in the Raman spectra.

The content of all the detected elements by AAS is presented in Table 1.

As can be seen in Table 1, regarding potential toxic elements, Hg (0.0012-0.0027 µg/g) was always found below the maximum allowed concentration level of 0.2 µg/g [1]. The content of Pb (0.80-9.0 µg/g) and Cu (3-13900 µg/g) was obtained for all the inks whilst Cr (3 µg/g), Ni (0.40 µg/g) and Cd (0.163 µg/g) were only above the technique's detection limit in the black ink. It was in this ink that the highest content of Pb was found (4.5 × the maximum allowed concentration level as defined by the ResAP(2008)1 [1]). According to the CoE's resolution, the maximum allowed concentration of Cu and Cr impurities in tattoos inks pertain to soluble copper and chromium VI. Although the content of these elements was above the maximum allowed levels in some of the analysed inks, none of the identified molecular compounds presented soluble copper or chromium VI. In fact, Raman spectroscopy only provided the identification of green and blue copper phtalocyanines in Forest is green and Tsunami blue inks, respectively, while the dioxazine violet compound was found in Violent violet (Fig. 2) and carbon black in the Black ink [11, 20]. Nevertheless, Cr and Ni are considered allergenic elements, and to minimize potential health risk for very sensitive individuals, its levels should not exceed 1 µg/g [21]. Furthermore, Pigment Violet 23, Pigment Blue 15, and Pigment Green 7 identified in violet, blue and green inks, respectively, are included in a list discouraging their use (Directive 76/768/EEC (Annex II) [22]) mentioned in the CoE ResAP(2008)1.

The Raman spectra of Deep yellow is shown in Fig. 3 along with the spectra of four different reference disazo pigments: Pigment Yellow 12, 126, 127 and 188. The high similarity of the pigments structures, makes it practically impossible to distinguish the four pigments by comparing their spectra. Nevertheless, similarly to the study of Bouchard *et al.* [23], Raman microscopy allowed the characterization of the Deep yellow ink pigment as belonging to the disazo group through specific bands such as the N=N symmetric stretching at 1399 cm$^{-1}$. Yet, Pigment Yellow 12 is the only of these pigments that is part of the negative list mentioned in the CoE ResAP(2008)1. The characteristic Raman bands from the monoazo Pigment Red 8 were identified in the red tattoo ink Fig.2. Many azo pigments employed in tattoos are not allowed for use in cosmetics because they may be decomposed in the skin by ultraviolet light yielding carcinogenic amines [7].

Conclusions and Outlook

A set of top brand tattoo inks was analysed by atomic and vibrational spectroscopic techniques to see if the composition of the inks matched the manufacturer's certification and if it complied with the CoE ResAp (2008)1 on requirements and criteria for the safety of tattoos and PMU. Three pigments present in the inks were identified as making part of the negative lists of the CoE resolution (pigments Violet 23, Blue 15 and Green 6). Besides, Ni and Pb levels were found above the maximum allowed concentrations of impurities in products for tattoos and PMU, and Cr and Ni were above the safe allergenic limit of 1 mg/g. Raman spectroscopy technique did not allowed to perceive any soluble Cu or Cr (VI) compounds. In future work, analysis by X-Ray Absorption Fine Structure spectroscopy (XAFS) will be carried out for Cr and Cu speciation and coordination number determination which, in the present study could have been useful in Tsunami blue, Forest is green, Violent Violet and Black inks due to their high total contents of these elements. Assessing the chemical state of such elements will provide extra information about the toxicity of the molecule to which the central atom is connected to. Nevertheless, Raman spectroscopy proved to be a suitable technique for the identification of most of the pigments present in the tattoo inks or at least the functional group. Alike in Bocca *et al.* [13] study, carbon black, Cu-phthalocyanine and $TiO_2$ were found in some of the Kuro Sumi inks. In their study they show that black tattoo inks contained nanoparticles (<100 nm) of carbon black whilst violet, blue and green inks presented nanoparticles of Cu-phthalocyanine and micro-sized particles containing titanium oxides. In future studies, an analysis of the size distribution of nanoparticles in Kuro Sumi inks should be verified. The results presented in this work show the importance of regulating tattoo ink composition. According to Cuyper *et al.* [7], uniform regulation at a global scale would positively contribute to the reduction of the risks and complications involved in the use of chemical compounds that may threaten the health of tattooed individuals, with special concern for heavy metals and carcinogenic aromatic amines. Moreover, better knowledge about substances used in the tattoo practice could be helpful to identify clinical problems occurring in medical practice.

## Acknowledgments

Mauro Guerra and Sofia Pessanha acknowledge the support of the Portuguese Foundation for Science and Technology for the grant BPD/92455/2013 and SFRH/BPD/94234/2013, respectively. This work was supported by the research centre grant no. UID/FIS/04559/2013 to LIBPhys-UNL, from the FCT/MCTES/PIDDAC. We thank HZB for the allocation of synchrotron radiation beamtime.

Table 1 Elemental concentrations ± standard deviation (μg/g) obtained by AAS in Kuro Sumi tattoo inks. The maximum allowed values are also displayed for comparison, as well as the detection limit (DL) and the quantification limit (QL) for AAS.

|  | Cr | Ni | Cu | Cd | Hg | Pb |
|---|---|---|---|---|---|---|
| Max. Allowed[1] | 0.2 (CrVI) | * | 25 (Cu soluble) | 0.2 | 0.2 | 2 |
| White | <DL | <DL | 3±1 | <DL | 0.0012±0.0001 | 3±1 |
| Black | 3.0±0.4 | 0.4±0.05 | 6.0±0.4 | 0.163<DL0.001 | 0.0202±0.0002 | 9.0±0.2 |
| Forest is green | <DL | <DL | 4400±200 | <DL | <DL | 0.80±0.01 |
| Tsunami blue | <DL | <DL | 13900±200 | <DL | <DL | 1.40±0.04 |
| Violent violet | <DL | <DL | 110±7 | <DL | 0.0022±0.0001 | 3.0±0.2 |
| Tomato red | <DL | <DL | 6±1 | <DL | 0.0027±0.0001 | 1.0±0.1 |
| Deep yellow | <DL | <DL | 6±1 | <DL | <DL | 0.80±0.04 |
| LD | 0.09 | 0.02 | 0.02 | 0.003 | 0.0005 | 0.02 |
| LQ | 0.26 | 0.06 | 0.07 | 0.005 | 0.0010 | 0.06 |

*as low as technically achievable

Figure Captions

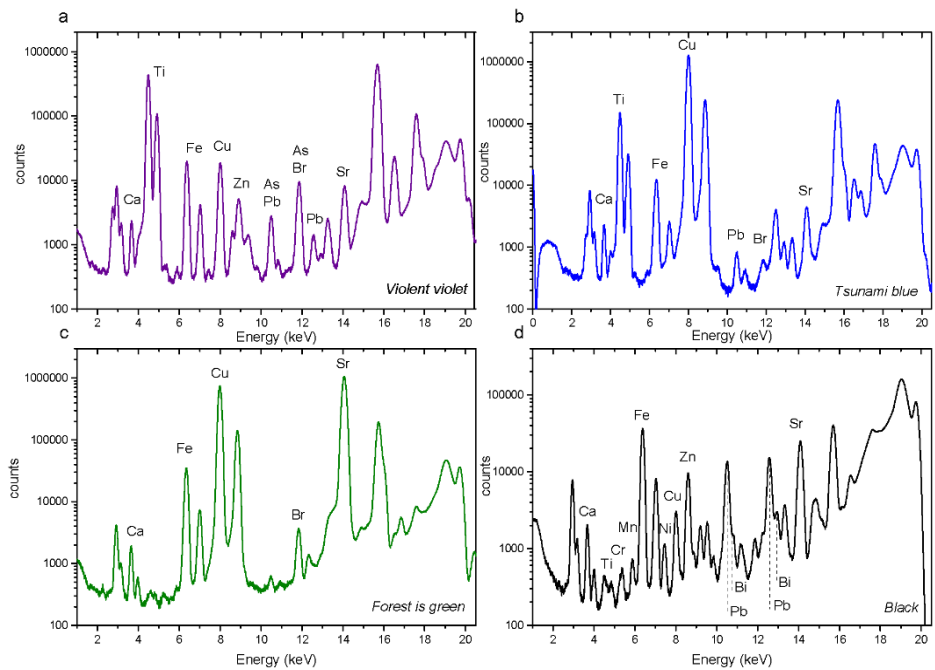

Fig. 1 Sy-XRF spectra obtained for 4 of the studied inks: a) violet, b) blue, c) green and d) black. X-ray characteristic peaks are identified.

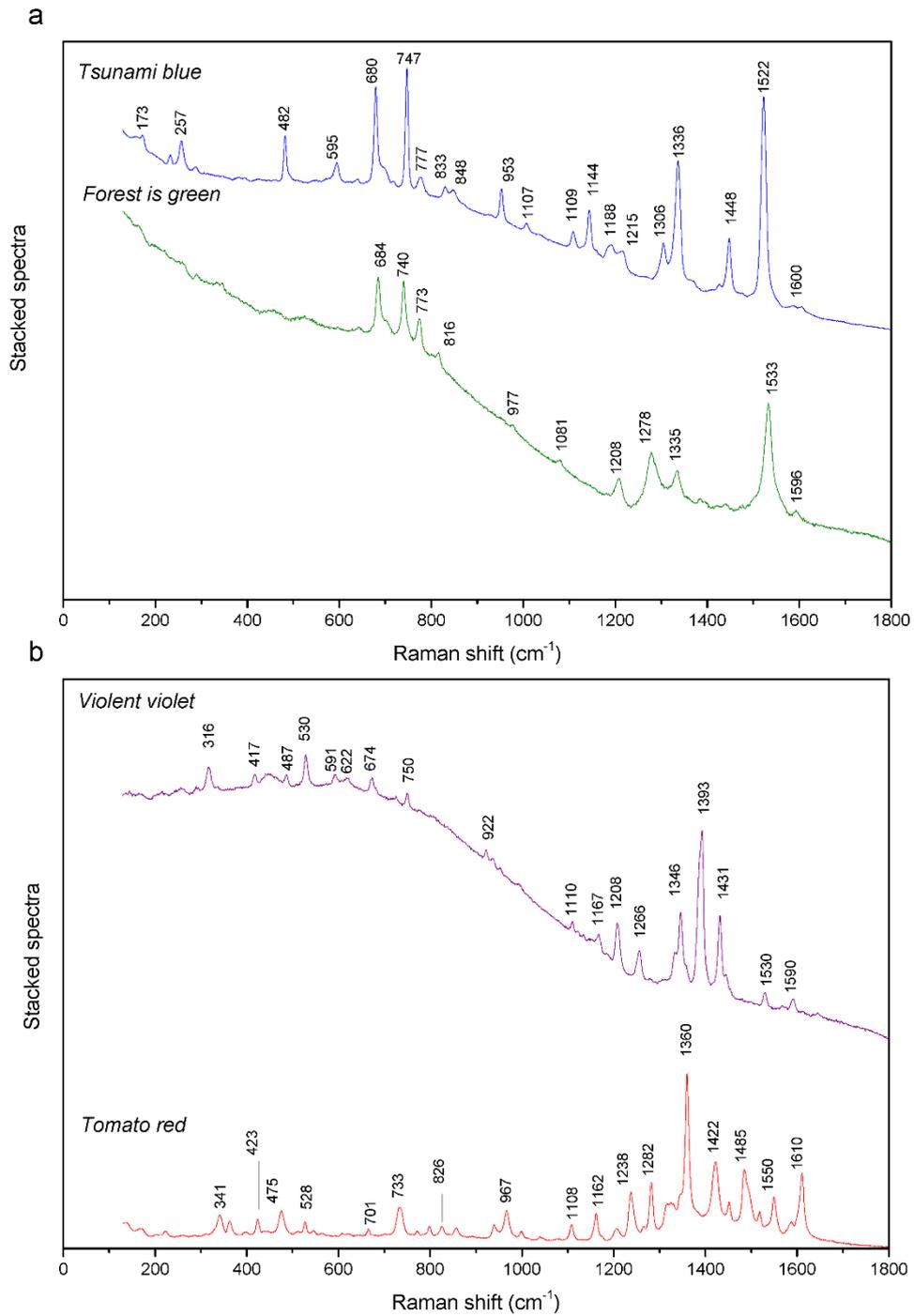

Fig. 2 Raman spectra obtained in a) blue and green and b) violet and red tattoo ink.

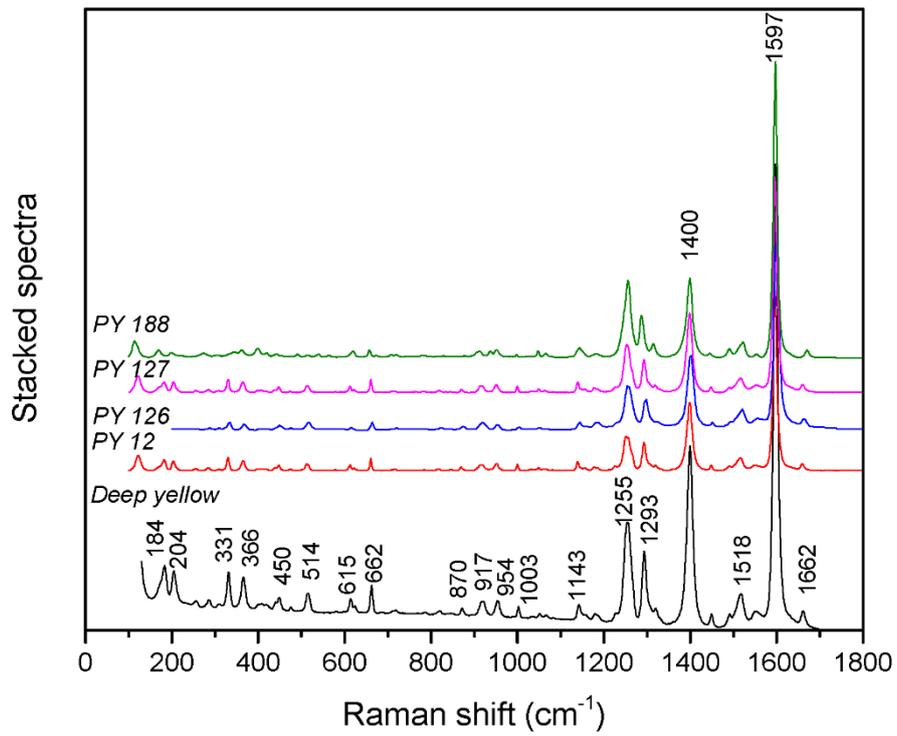

Fig. 3 Comparison of Raman spectrum obtained from Deep yellow ink with Pigment Yellow 12, 126, 127 and 128 reference spectra [18].